\documentclass[review]{elsarticle}

\usepackage{hyperref}
\usepackage{xcolor}

\usepackage{graphicx}
\usepackage{tikz}
\usepackage{callouts}
\usepackage{amsfonts}
\usepackage{mathtools}
\usepackage{amsthm}
\usepackage{bbm}

\usepackage{multirow}
\usepackage{booktabs}
\usepackage{tabularx}
\usepackage[graphicx]{realboxes}
\usepackage{float}

\usepackage{makecell}

\usepackage{amsmath}

\def\vd{{\boldsymbol{d}}}

\def\vh{{\boldsymbol{h}}}

\def\vt{{\boldsymbol{t}}}

\def\vx{{\boldsymbol{x}}}
\def\vy{{\boldsymbol{y}}}
\def\vz{{\boldsymbol{z}}}

\def\mA{{\boldsymbol{A}}}

\def\mE{{\boldsymbol{E}}}

\def\mH{{\boldsymbol{H}}}

\def\mK{{\boldsymbol{K}}}

\def\mQ{{\boldsymbol{Q}}}

\def\mV{{\boldsymbol{V}}}
\def\mW{{\boldsymbol{W}}}
\def\mX{{\boldsymbol{X}}}

\def\gE{{\mathcal{E}}}

\def\gG{{\mathcal{G}}}

\def\gL{{\mathcal{L}}}

\def\gN{{\mathcal{N}}}

\def\gS{{\mathcal{S}}}

\def\gV{{\mathcal{V}}}

\newcommand{\R}{\mathbb{R}}

\newcommand{\eg}{\textit{e}.\textit{g}., }
\newcommand{\ie}{\textit{i}.\textit{e}., }

\newcounter{bxincomm}
\definecolor{aqua}{rgb}{0.00,0.67,0.80}

\newcounter{xycomm}

\newcounter{ygcounter}

\newcommand{\ygc}[1]{\ygc{\stepcounter{ygcounter}{\bf [YG's comment \arabic{ygcounter}: #1]}\;}}

\journal{Journal of \LaTeX\ Templates}

\begin{document}

\begin{frontmatter}

\title{Graph Representation Learning for \\ Interactive Biomolecule Systems}


\author[mymainaddress]{Xinye Xiong}

\author[mymainaddress,mysecondaryaddress]{Bingxin Zhou\corref{mycorrespondingauthor}}
\cortext[mycorrespondingauthor]{Corresponding author}
\ead{bingxin.zhou@sjtu.edu.cn}

\author[mymainaddress,mysecondaryaddress]{Yu Guang Wang}

\address[mymainaddress]{Institute of Natural Sciences, Shanghai Jiao Tong University, Shanghai, 200240, China}
\address[mysecondaryaddress]{Shanghai National Center for Applied Mathematics (SJTU Center), Shanghai, 200240, China}

\begin{abstract}
Advances in deep learning models have revolutionized the study of biomolecule systems and their mechanisms. Graph representation learning, in particular, is important for accurately capturing the geometric information of biomolecules at different levels. This paper presents a comprehensive review of the methodologies used to represent biological molecules and systems as computer-recognizable objects, such as sequences, graphs, and surfaces. Moreover, it examines how geometric deep learning models, with an emphasis on graph-based techniques, can analyze biomolecule data to enable drug discovery, protein characterization, and biological system analysis. The study concludes with an overview of the current state of the field, highlighting the challenges that exist and the potential future research directions.
\end{abstract}


\end{frontmatter}

\section{Introduction}
Studying molecular mechanisms and interactions at different levels of the biological systems is fundamental for understanding organisms, discovering disease pathways, developing novel drugs, and designing functional proteins like enzymes and antibodies. The emergence of deep learning techniques has facilitated the automatic analysis of biological entities by transforming large datasets into tractable numerical representations for trainable algorithms. 

One prevalent strategy for capturing the interactions among subunits, \eg amino acids in a protein or atoms in a chemical compound, is to construct a graph where nodes denote the subunits, and edges indicate the pair-wise interactions between them. For example, a molecule graph represents atoms as nodes connected by chemical bonds. Similarly, a protein graph often employs amino acids as nodes, while edges can reveal sequential neighborhoods and/or spatial proximity. Moreover, graphs can depict higher-level interactions, such as gene expression and metabolic pathway prediction, where a graph captures information exchange between different groups or systems, and nodes stand for molecules, cells, or more complex entities.

Proteins are essential biomolecules with diverse functions in living organisms, including transport, storage, membrane composition, and enzymatic activity. To perform these functions, proteins need to interact physically with small molecules, DNA/RNA, and other proteins. Understanding the function of proteins is crucial for gaining biological insights into protein engineering for disease treatment and prevention, among other life sciences applications.

Protein function is determined by its structure, which has been studied through techniques such as X-ray crystallography, NMR spectroscopy, and Cryo-EM since the 1960s. With the increasing availability of protein structures, particularly with the emergence of \textit{in silico} folding methods \cite{baek2021accurate,chi2016selection,jumper2021highly, varadi2022alphafold}, graphs have become a popular tool for characterizing the spatial organization of proteins. A graph is a two-dimensional representation of structured data that encodes pairwise topology between node objects, such as atomic connectivity through chemical bonds.

One significant breakthrough in AI methods for predicting protein structure comes with \textsc{AlphaFold 2} \cite{jumper2021highly}. This algorithm achieves precision comparable to wet lab experiments by combining \textsc{Transformer} and structure learning models to extract general protein features from large protein sequences. Another notable method is \textsc{RosettaFold} \cite{baek2021accurate} with SE(3) equivariant neural networks and achieves slightly lower performance. Both methods utilize geometric deep learning as a key component of their success. Since then, structure deep learning has been a tool for protein-related prediction tasks, \eg protein folding and inverse folding, directed evolution, protein-protein docking, protein-small molecule docking, and \textit{de novo} protein sequence generation. These advances are revolutionizing the traditional pharmaceutical industry, which searches for target protein-based drugs by enumerating drug-like molecules on high-throughput screening (HTS) technologies at a high cost of time and labor \cite{du2022molgensurvey}.

The remainder of the paper is organized as follows: Sections~\ref{sec:4homoHeteroGraphs} and \ref{sec:moleculeCharacterization} introduce the learning representation of biological systems and micro- and macromolecules. Section~\ref{sec:GRL} reviews typical graph machine learning tasks on biomolecules. Section~\ref{sec:representation learning} reviews a variety of representation learning methods for extracting and processing biomolecule data and accomplishing downstream tasks. Section~\ref{sec:application} reviews various applications of geometric representation learning in drug discovery, protein understanding, and biological system analysis. Section~\ref{sec:benchmark} summarizes popular benchmark datasets on protein complexes, protein-ligand binding, antibody-antigen structures, and knowledge graphs. Section~\ref{sec:discussion} discusses and concludes the paper with industrial applications.

\section{Geometric Representation in Biological Systems}
\label{sec:4homoHeteroGraphs}
\begin{figure}
    \centering
    \includegraphics[width=\textwidth]{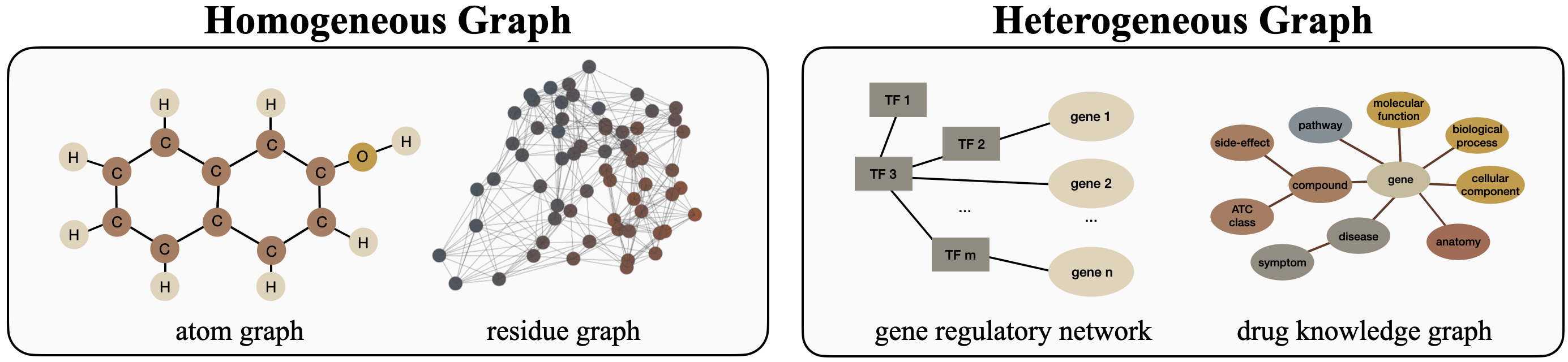}
    \vspace{-8mm}
    \caption{Examples of homogeneous and heterogeneous graphs.}
    \label{fig:graphs2types}
\end{figure}

Interactions play a crucial role in depicting the complexity of biological systems, and they can be represented as graphs that are amenable to computational analysis. The nature and scale of the interactions dictate the categorization of the graphs into four distinct types, as illustrated in Figure~\ref{fig:graphs2types}.

\subsection{Homogeneous Graph}
When representing structured entities using graphs, it is customary to utilize homogeneous graphs, in which all nodes and edges are of the same type. In the field of structural biology, such homogeneous graphs are commonly employed to depict chemical compounds, proteins, or peptide chains, as we will discuss presently.

\subsubsection{Atom Graphs for Chemical Compounds}
In the realm of computational biology, chemical compounds are the smallest unit of interest in drug design. Typically, drug candidates are small molecules composed of several tens or hundreds of atoms. The proliferation of high-throughput screening technologies and the advent of computer technology has paved the way for the computational analysis of bioactive drug-like molecules. To this end, it is necessary to represent these compounds using syntactic structures that can be processed by algorithms. One widely used representation is the Simplified Molecular Input Line Entry System (SMILES) \cite{weininger1988smiles}, which encodes isomeric information into a one-dimensional sequence. While SMILES does not encode stereochemistry, an alternative choice is the \emph{molecular graph} representation, which represents a compound as a graph, with atoms as nodes and covalent bonds as edges between pairs of nodes.

The utilization of structural information representation has been extensively employed in the field of drug discovery, particularly in drug-target interaction (DTI) \cite{abbasi2021deep,zhao2021biomedical,du2016insights}, binding affinity prediction \cite{shen2020machine,bagherian2021machine}, and drug generation \cite{stephenson2019survey,jimenez2020drug}. Depending on the specific formulation of the learning task, it can be a learning task at the node level (\eg determine which nodes of the target drug will interface with the pocket) or the graph level (\eg forecast whether the drug will bind to the protein with a given pocket). A comprehensive discourse of the prevalent learning tasks will be expounded in Section~\ref{sec:GRL}.

\subsubsection{Residue Graphs for Proteins}
Proteins perform diverse functions in living systems, such as transport, storage, membrane composition, and enzymatic action. The function of a protein is determined by its unique structure, which serves as a natural foundation for incorporating structural information into algorithms for protein analysis. Graphs have become a popular tool for characterizing protein spatial arrangements. In a residue graph, nodes usually represent amino acids, and edges are constructed to express the chemical or geometrical relationships between node pairs. Alternatively, nodes can be defined as heavy atoms on the main chain to capture more detailed information at finer scales. For a comprehensive discussion on protein graph construction, we recommend readers to \cite{fasoulis2021graph,ovchinnikov2021structure}.

Predicting unseen interactions between a protein and one or more other biological entities has caught great attention in recent studies. One such application is vaccine design, which studies to identify effective interactions between antigens and antibodies to  combat pathogenic activity \cite{graves2020review,wilman2022machine,ruffolo2022antibody}. RNA-protein interaction (RPI) identifies the binding sites and binding affinities between ribonucleic acid (RNA) and RNA binding proteins to understand biological processes such as post-transcriptional regulation of genes \cite{RBP2019pan,yan2020review,wei2022protein,cui2022protein}. Constructing the quaternary structure of a protein complex is another vital perspective to understanding the protein functions in the living system, which also requires examining the protein-protein interactions \cite{susanty2021review,soleymani2022protein}.

\subsection{Heterogeneous Network}
Heterogeneous graphs are frequently used to represent integrated networks that depict relationships among diverse entities such as gene expression data and drug-disease association networks. These graphs differ from homogeneous graphs in that they comprise various node and/or edge types, providing a wealth of information and multiple perspectives for the underlying system.

\subsubsection{Molecular Interaction Network}
A biological system, whether it be a living organism or a human body, is comprised of numerous entities that interact with each other to form a complex network. Nodes in these networks may represent genes, enzymes, proteins, metabolites, other biomolecules, or environmental factors, depending on the network's properties \cite{liu2020computational}. In the context of graph representations, different nodes are connected by (directed or undirected) edges that indicate their interactions with each other. For example, a gene regulatory network \cite{muzio2021biological} might feature a directed edge from gene A to gene B, to imply that gene A directly regulates the expression of gene B without the intervention of other genes. Similarly, a metabolism network \cite{lawson2021machine} studies enzymatic reactions (edges) that catalyze one metabolite (node) to another. 

The prediction of edge connections, which represent the interactions between molecules, is often the primary goal of graph analysis. In gene regulatory networks, interpreting links between unconnected nodes from gene expression time series data provides insight into the mechanistic regulation of gene expression and allows researchers to infer the phenotype from observed events, given that not all regulatory relationships are adequately captured. In metabolism networks, predicting the metabolic pathway or the pathway dynamics of a specific compound is essential for identifying key metabolites and associated genes and prioritizing potential targets for metabolic engineering.

\subsubsection{Drug Knowledge Graph}
The prediction of interactions between drugs and targets is crucial for understanding the mechanisms of drug action and identifying potential adverse effects, thus facilitating cost-efficient and safe drug discovery. Network-based approaches usually take existing drugs and diseases as a big interactive network to interpret drug-target associations and propagate the target drug. Biomedical knowledge graphs provide rich and diverse information about various entities and their relationships, such as drug-target interactions and semantic relations between drug entities, to facilitate the interpretation of unknown relationships \cite{lotfi2018review,pan2022deep}. Adverse drug reaction prediction is also important, as unexpected interactions between drugs can lead to polypharmacy side effects that undermine the intended therapeutic effects. Accurate prediction of drug-drug interactions can thus increase the success rate of clinical trials and reduce costs in drug discovery.
Interpreting the unknown relationships of given entities also plays a key role in adverse drug reaction prediction, which labels undesirable interactions of drugs and targets that fail to meet the intended therapeutic requirements. For instance, a common factor that affects phase III clinical trials is polypharmacy side effects raised by unexpected interactions between drugs \cite{ventura2010adverse}. Therefore, effective prediction of drug-drug interactions is substantial for sparing the cost and increasing the success rate of clinical trials for drug discovery.

\section{Characterization of Molecules}
\label{sec:moleculeCharacterization}
In the context of deep learning, numerical representations of biological entities such as atoms, compounds, proteins, genes, or cells are required. To set up the problem, different descriptions of input biomolecules are used to reflect the preferred inductive bias, which is closely related to the selection of deep learning techniques. For instance, protein sequences are typically processed with language models to extract contextual representations along one-dimensional chains, allowing for capturing long-range relationships between different atoms or amino acids along the same chain. On the other hand, grid-like representations are handled by convolutional networks, where atoms or amino acids within smaller receptive fields are more influential to the central instance. This section provides a summary of the common rules for characterizing molecules.

\subsection{Node Types}
As mentioned earlier, biomolecules possess a well-defined hierarchy. For example, proteins consist of amino acids, which in turn are composed of multiple atoms. In this section, we will discuss the construction of a biological system in three hierarchies: atoms, amino acids, and macromolecules.

\paragraph{Atom}
A chemical entity comprises a set of atoms that are interconnected by means of chemical bonds. The atomic characteristics dictate the shape and properties of the molecule, rendering it sensible to represent a chemical compound with its constituent atomic symbols, such as C for carbon, O for oxygen, and N for nitrogen. The description of atoms can be supplemented by their physicochemical properties, such as the charges of atoms and the number of attached hydrogen atoms \cite{wang2021protein}. The shape or conformation of the molecule is crucial for studying its functionality. For small molecules, the torsion angles (or dihedral angles) of four consecutive covalently bonded atoms are usually used.

\paragraph{Amino Acid}
The primary structure of a protein refers to the specific arrangement of amino acids that make up its structural foundation. Amino acids, excluding those that are uncommon or non-traditional, can be classified into $20$ different groups. Each type of amino acid is commonly designated by a unique one-letter code, such as R for Arginine and K for Lysine, and these codes correspond to various characteristics including hydrophobicity, polarizability, and net charge. For an extensive overview of the diverse properties of amino acids, please refer to \cite{soleymani2022protein}.

\paragraph{Molecule}
When investigating the intermolecular interactions within a biological system or between different systems, it is commonplace to model molecules and their interactions using nodes and edges in a graph \cite{jeong2000large,ma2003connectivity}. The choice of nodes depends on the particular system under investigation. For example, in gene regulatory networks, nodes could represent transcription factors and target genes \cite{wang2020inductive}. In metabolic networks, nodes might be metabolites-reactions \cite{beguerisse2018flux} or enzyme-substrate pairs \cite{kroll2021deep}. Alternatively, in drug knowledge graphs, nodes may be disease, gene, compound, protein, and other relevant entities \cite{bonner2022review,zeng2022toward}. 

\subsection{Molecule Modalities}
There exist various modalities to represent a given biomolecule. For instance, a protein's chemical composition can be expressed using its amino acid sequence, and its folded tertiary structure can be used to describe the inter-residue spatial relationships. We provide a brief summary of some commonly used methods for representing biomolecules.

\begin{figure}[t]
    \centering
    \includegraphics[width=\linewidth]{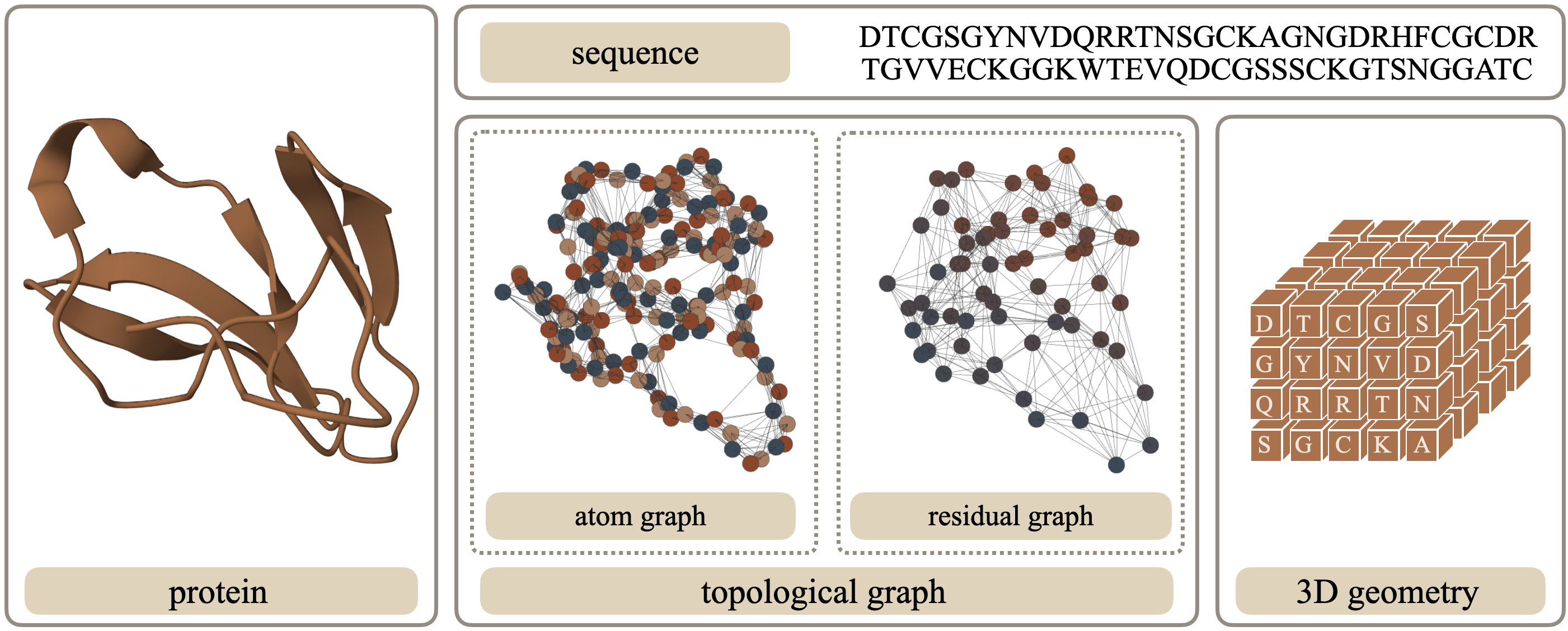}\vspace{-8mm}
    \caption{Different modalities of a molecule. A protein can be represented numerically in different dimensions. See Section~\ref{sec:moleculeCharacterization} for more discussion.}
    \label{fig:molReps}
\end{figure}

\paragraph{1D Sequence}
As previously discussed, one approach for representing a molecule is through character-level tokenization \cite{OFER20211750}, where each component (\eg an atom or amino acid) is represented by a character in a string. For example, complex compounds can be described using SMILES \cite{weininger1988smiles} or self-referencing embedded strings (SELFIES) \cite{krenn2020self}, while the primary structure of a protein can be encoded as a unique sentence composed of $20$ different amino acid types. In proteins, amino acids, motifs, and domains are comparable to words, phrases, and sentences in human languages, where sequential representations are typically encoded using language models such as 1D CNNs and attention mechanisms.

\paragraph{2D Graph}
While the sequential representation of a molecule provides a detailed breakdown of its components, it fails to consider the intramolecular steric interactions between them, such as the force fields between atoms. Graph data presents an alternative way to describe a molecule's geometric connectivity by reducing its conformation without explicitly defining a 3D coordinate system. In essence, a graph is composed of a set of nodes that represent the characters in the 1D sequence (atoms or residues), with edges connecting nodes based on identified relationships. These relationships can be either explicitly defined by biological connections (\eg chemical bonds) \cite{lim2019predicting,ganea2021geomol,xu2021geodiff} or implicitly defined by geometric connections (\eg pair-wise Euclidean distances in 3D space) \cite{jin2021iterative,zhou2022lgn,chen2022dproq}. The edge connectivity is stored in an adjacency matrix, which is critical for constructing aggregation rules that establish node representations. However, constructing an appropriate adjacency matrix can be challenging, particularly for large molecules like proteins. Sparse connectivity may fail to capture the complete geometry of the underlying topological manifold, while an overly dense adjacency matrix may impair computational efficiency and potentially dilute valuable information for learning tasks.

\paragraph{3D Geometry}
The spatial relationships between internal entities (\eg amino acids or non-hydrogen atoms) in molecular structures can be represented in 3D space by their 3D coordinates. There are several variants of 3D geometric representation. If neglecting the pair-wise connections, the group of entities is called a point cloud \cite{eismannprotein,tubiana2022scannet}; if the nodes are assumed aligning to the grid, it can be viewed as a 3D box \cite{derevyanko20183dcnn}; if a molecule's shape is a major concern over its internal structure, a 3D mesh can be leveraged to describe the surface \cite{gainza2020deciphering,somnath2021multi}; otherwise, if the nodes are connected discretely in the space, we take it as a 3D graph \cite{zhou2022lgn}. 
The choice of representation depends on the specific properties of interest, and all of them require consideration of rotation and translation equivariance in 3D space by geometric deep learning \cite{bronstein2017geometric,bronstein2021geometric}.

\subsection{Feature Preparation}
The features of nodes and edges in a network are informed by domain-specific knowledge, which may exert a significant influence on learning outcomes. Some widely-used input functions are enumerated below.

\paragraph{Atom Feature} 
An in-depth examination of proteins at the atomic level provides a more nuanced understanding, as the energy of inter-residue interactions is influenced by the number of atoms in contact \cite{mirny2001protein}. These atoms can be identified by their specific types, with some common examples including N, C, O, and S \cite{jiang2022predicting}. Atom types also exhibit various physicochemical attributes, such as charge (positive, negative, or neutral) \cite{politzer1970properties}, polarity (polar or non-polar) \cite{laidig1990properties}, and aromaticity (aliphatic, aromatic, or neutral) \cite{baldridge2000silabenzenes}, which augment their distinctive characteristics. In addition, researchers have defined spatial attributes for atom nodes, such as the number of attached non-hydrogen or hetero atoms representing heavy-valence or hetero-valence properties \cite{wang2021protein, son2021development}. Rather than hand-crafting atom-level feature vectors, an alternative is to employ fingerprints, a type of sparse descriptor containing thousands of values, to represent the structural and physicochemical properties of atoms \cite{zhao2021biomedical}.

\paragraph{Residue Feature}
When considering an amino acid within a biomolecule, such as a protein, its physicochemical and geometric characteristics play a critical role in determining its potential for interaction within the molecule structure. A protein is typically represented by one-hot encoding the \emph{residue types} to $20$ or $26$ dimensions \cite{soleymani2022protein}. The rigidity, flexibility, and internal motion of the residue can be identified by the \emph{B-factor}, which is determined through crystallization processes \cite{sun2019utility}. Geometric properties, such as the \emph{3D position} of the residue, are typically represented by its $\alpha$-carbon (or C$\alpha$) coordinates \cite{correa1990building}, while the \emph{solvent-accessible surface area} (SASA) is used to identify whether the residue is on the protein's surface or in the core \cite{moret2007amino}. The relative positions of amino acids in the protein chain can be expressed using trigonometric values of \emph{backbone dihedral angles} $\{\sin, \cos\} \circ \{\phi, \psi\}$ derived from the positions of the backbone atoms N, C$\alpha$ and C \cite{gogoi2023protein}. In addition to scalar features, vector features can also be informative, such as the direction of C$\beta$-C$\alpha$ \cite{jumper2021highly,dauparas2022proteinMPNN}. Incorporating such directional attributes requires designing specialized propagation rules, such as graph vector perceptrons (GVP) \cite{jing2020learning}.


\paragraph{Edge Feature}
The attributes of edges describe the relationship between interconnected nodes. The \emph{inter-residue distances} of adjacent nodes can be derived directly using their Euclidean distance in 3D space. Additionally, Gaussian radial basis functions (RBF) can be used to project scalar distance into higher dimensions for better expressivity \cite{ganea2021independent,ingraham2019generative}. \emph{Relative positional encoding} represents the sequential distance of closely located residues on the amino acid chain \cite{jumper2021highly,dauparas2022proteinMPNN,liu2022rotamer}. The positions of heavy atoms in two connected residues encode local interactions with fine-grained \emph{local frame orientation} \cite{stark2022equibind,zhou2022lgn}.

\begin{figure}[t]
    \centering
    \includegraphics[width=\textwidth]{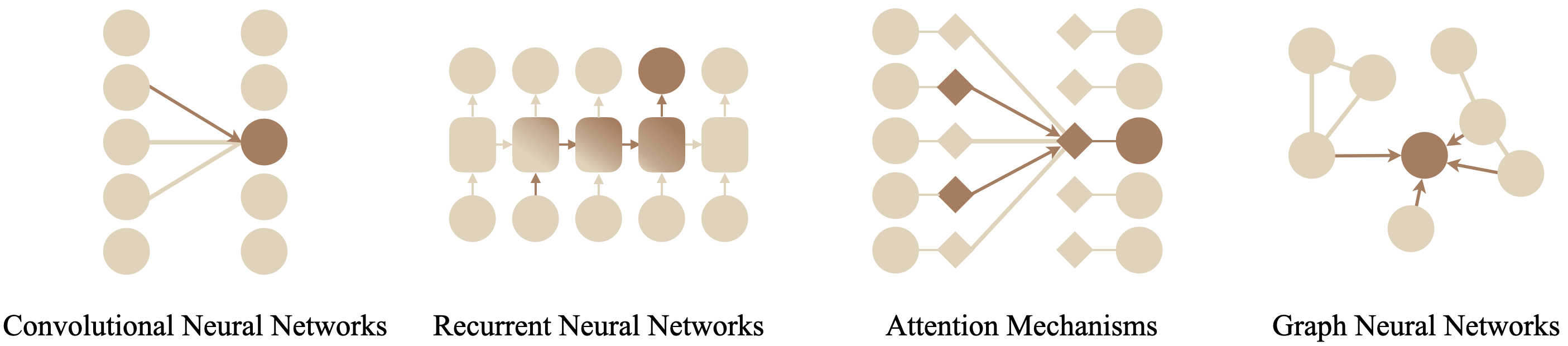}\vspace{-8mm}
    \caption{The characterization of biomolecules and the corresponding deep learning models embeds inductive biases of molecules.}
    \label{fig:inductiveBias}
\end{figure}

\section{Geometric Learning Tasks}
\label{sec:GRL}
A graph $\gG=(\gV,\gE)$ can describe the structure of a given molecule with a set of $n=|\gV|$ nodes and edges $\gE$, where $e_{ij}\neq0$ indicates node $v_i$ is connected to $v_j$ and $e_{ij}=0$ otherwise. For computational convenience, the pair-wise graph connection is stored in an adjacency matrix $\mA\in\R^{n\times n}$, which is a symmetric matrix for undirected graph, \ie, $e_{ij}=e_{ji}$. The numerical representation for nodes and edges, if applicable, are respectively described by $\mX$ and $\mE$.

\subsection{Graph Representation Learning}
In graph representation learning, the main objective is to learn a powerful and informative latent representation of the graph or nodes, which can be further utilized in various downstream applications such as node- or graph-level attribute predictions.

\paragraph{Discriminative Learning}
Supervised or semi-supervised discriminative learning models are usually applied to predict the optimal labels from input $\mX$ using large sets of labeled data. The discriminative models essentially learn to estimate the probability distribution $P(\vy|\mX)$, where $\vy$ represents the target variable specific to the task at hand.

\paragraph{Generative Learning}
Despite the considerable progress achieved by supervised learning models in predicting labels, their reliance on extensive manual labeling poses challenges such as generalization errors, spurious correlations, and adversarial attacks. In computational biology, the limited availability of labeled data necessitates the development of approaches that can learn from fewer manual labels, samples, and trials. In some cases, the goal of learning is to generate new molecule samples $P(\mX|\vy)$ from the joint distribution $P(\mX,\vy)$ with specified $\vy$.

\begin{figure}
    \centering
    \includegraphics[width=\textwidth]{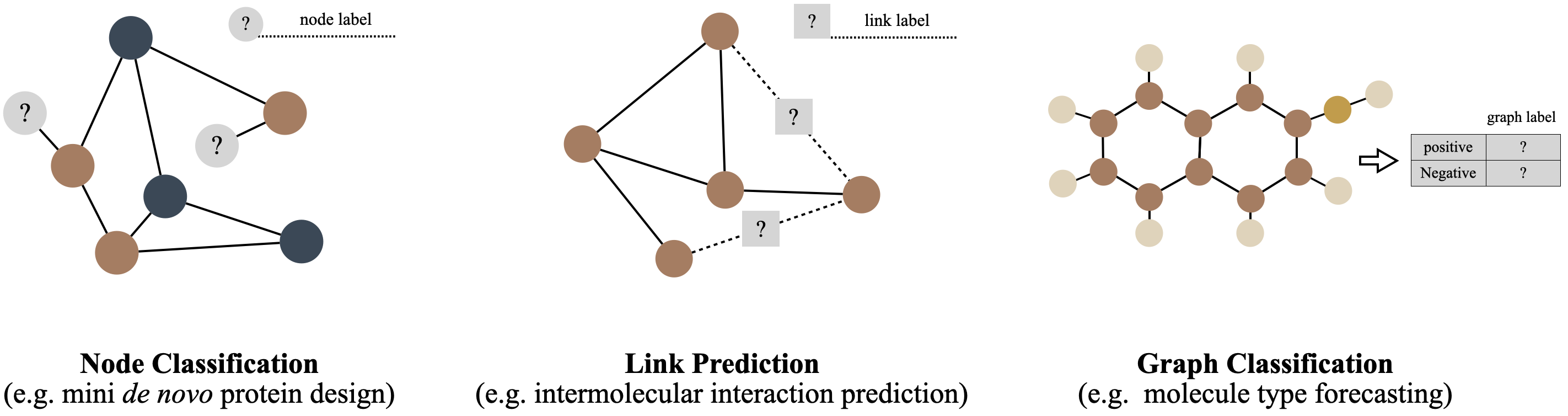}\vspace{-8mm}
    \caption{Three types of downstream tasks in geometric deep learning at node, edge, and graph levels.}
    \label{fig:GRLtask}
\end{figure}

\subsection{Downstream Tasks}
Graph data is composed of nodes, edges, and the graph itself. Typically, representation learning is done at the node level, but the resulting vector representation can be utilized for all three levels of label assignment tasks.

\paragraph{Node-level Properties Prediction}
At the node level, graph convolutional layers encode the topology of an attributed graph by creating a matrix $\mH\in\R^{n\times h}$, where each of the $n$ nodes is represented by an $h$-dimensional vector that aggregates information about its neighborhood. Then, readout layers assign each node $\gV_i\in\gV$ a label $y_i=f(\vh_i)$. 
The learning tasks are named node classification or regression, depending on whether the label $y_i$ is discrete or continuous. For example, identifying the type of an amino acid in a protein is a \emph{node classification} task, while predicting the SASA of a node is a \emph{node regression} task.

\paragraph{Link Prediction}
The process of learning edge properties is similar to node-level learning, except that the subjects of labeling become edges instead of nodes. Typically, a label is a binary indicator for the existence or non-existence of a connection between nodes or a decimal weight indicating the strength of the relationship. The objective of \emph{relationship inference} or \emph{link prediction} is to estimate the pairwise connections of nodes, represented as $y_{ij} = f(\vh_i, \vh_j)$, given a set of node representations $\mH$. The concatenation of hidden representations of node pairs is used to predict whether there exists a connection between $\gV_i$ and $\gV_j$ through binary classification. Link prediction tasks are commonly used in analyzing heterophilic graphs, such as molecular interaction networks \cite{jiang2021learning} and drug repurposing graphs \cite{sadeghi2019analytical}.

\paragraph{Graph-level Properties Prediction}
Graph-level tasks refer to the prediction of labels for a collection of graphs, denoted as ${\gG_1,\dots,\gG_N}$, with corresponding labels ${y_1,\dots,y_N}$. Graphs can differ in size and structure, in contrast to image classification where the number of pixels is consistent across different images. For example, in \textbf{D\&D} \citep{dobson2003distinguishing,shervashidze2011weisfeiler} that classify proteins into enzyme and non-enzyme structures, a protein structure can consist of dozens to over $5,000$ amino acids. Developing graph pooling strategies that can unify varying-sized graphs into a low-dimensional representation is beneficial for these tasks. Moreover, when predicting interactions between pairs of graphs, concatenation layers may be necessary to combine the representations of different graphs.

\section{Representation Learning Strategies}
\label{sec:representation learning}
This section discusses three main approaches to establishing a learning pipeline for biomolecules. We will begin with basic encoding modules for sequential or spatial molecular representation learning. Since encoding strategies can be combined in parallel or subsequently, we will also provide examples of multimodal inputs that offer more than one modality or scale of biomolecules. Given that labeled biomolecule datasets are limited in practice, we will summarize some essential principles for model augmentation. While we will use residue graphs or protein sequences as examples for explaining the models, they can be easily adapted or generalized to other types of molecules such as chemical compounds.

\subsection{Topological Encoding}
We present three commonly used neural networks in the field of molecular learning: graph networks, recurrent networks, and attention-based networks. Although not all introduced modules were initially designed for graph inputs, they can be used jointly or subsequently to fulfill various roles in a learning pipeline. For instance, \textsc{IEGMN} \cite{ganea2021independent,stark2022equibind} utilizes attention mechanisms in designing equivariant message passing for residue graphs; \textsc{GVP-GNN} \cite{jing2020learning,jing2021equivariant} designs attention-like vector perceptron to transmit both scalar and vector nature across graph nodes; \textsc{DeepFRI} \cite{gligorijevic2021structure} learns LSTM-extracted residue-level features with GCN.

\subsubsection{Neural Message Passing}
As discussed in Section~\ref{sec:moleculeCharacterization}, the topology of proteins is crucial for understanding their functionality. To characterize protein data, various inductive biases can be incorporated, such as 3D density mappings with CNNs \cite{torng20173d,francoeur2020three,lu2022machine} and interactive propagation with GNNs \cite{dauparas2022proteinMPNN}. The latter is typically formulated under the framework of \emph{(neural) message passing} \cite{gilmer2017neural}. For a biomolecular graph, the hidden representation $\mH_i^{(\ell+1)}$ for the $i$th node in the $(\ell+1)$th layer is updated by
\begin{equation} \label{eq:mpnn}
    \mH_i^{(\ell+1)} = \gamma\left(\mH^{(\ell)},\square_{\gV_j\in \mathcal{N}(i)}\phi(\mH_i^{(\ell)},\mH_j^{(\ell)},\mA_{j i})\right),
\end{equation}
where the information from all one-hop neighbors $\gV_j\in\gN(i)$ in $\mH^{(\ell)}$ is aggregated using a differentiable and permutation-invariant function $\square$, such as summation, average, or maximization. The $\gamma,\phi$ represent another two differentiable functions for transformation.

As a protein's atomic dynamics remain unchanged regardless of its translation or rotation within a system \citep{han2022geometrically}, some propagation rules incorporate the inductive bias of symmetry into structure-based models, requiring the learned 3D protein transformations to belong to the Special Euclidean group SE(3) \cite{bronstein2021geometric,marsden2013introduction}, which is the set of rotation and translation equivariant transformations in 3D space, \ie $(\mathcal{R},\vt)\mX=\mathcal{R}\mX+\vt$ with some rotation matrix $\mathcal{R}\in\text{SO(3)}$ and translation vector $\vt\in\R^3$. These algorithms avoid expensive data augmentation and typically need the propagation to consider positional encoding of 3D coordinates $\mX^{\prime}\in\R^{n\times 3}$ \cite{zhou2022lgn,yi2022approximate,satorras2021n}.

\subsubsection{Autoregressive Sequential Propagation}
Sequence-based inferences frequently assume a conditional relationship for $P(\vy) = \prod_{i=1}^n P(\vy_i|\vy_1,\dots,\vy_{i-1})$. In \textit{de novo} protein design, autoregressive methods are used to predict the next masked amino acid token based on previously inferred amino acids \cite{alley2019unified, biswas2021low, shin2021protein}. 
Such an assumption satisfies the memory embedding criteria of recurrent neural networks $R(\cdot)$, which joins the historical embeddings and current observation $\mX^{(t)}$ to determine its hidden representation:
\begin{equation} \label{eq:rnn}
    h(\mX^{(t)})= R\bigl(h(\mX^{(t-1)}), z(\mX^{(t)})\bigr) = \sigma\bigl(\mW_1 z(\mX^{(t)})+\mW_2 h(\mX^{(t-1)})+\boldsymbol{b}\bigr).
\end{equation}
Since the vanilla recurrent mechanism dampens the importance of early encoded states, gating mechanisms \cite{hochreiter1997long} introduce additional memory cells to guide the transmission of useful memory in previous steps. Gated recurrent networks have been widely applied in time series analysis for directional relationship inference. Because sequence representations of proteins neither specify an origin or termination nor a query direction in nature, BiLSTM \cite{schuster1997bidirectional}
becomes a popular choice due to its enhanced performance in forward and reverse long-short-term memory encoding \cite{bepler2018learning,hu2019improved,shi2022dream}.

\subsubsection{Pair-wise Relationship by Attention Mechanism}
Inferring connections between tokens (\ie amino acids) in either direction along the protein sequence using a recurrent network may miss important relationships between tokens that are sequentially far apart but spatially close. Attention mechanisms provide an alternative by considering all pair-wise relationships in a protein to infer the most comprehensive interactions between contact tokens. \textsc{Transformer} \cite{vaswani2017attention} as an advanced self-attention mechanism, has gained popularity due to its ability to capture global dependencies among input instances. \textsc{Transformer}, an advance in self-attention, has seen a surge in recent breakthroughs. For instance, \textsc{AlphaFold 2} \cite{jumper2021highly} leverages evolutionary Transformers (\textsc{Evoformer}) for pair-wise residue representations, and \textsc{ESM-if1} \cite{hsu2022learning} applies \textsc{Transformers} on scalar and vector features of residues. Fundamentally, a self-attention layer captures global dependencies among input instances $\mX$ by
\begin{equation} \label{eq:attn}
    \operatorname{Attention}(\mQ, \mK, \mV)=\operatorname{softmax}\left({\mQ \mK^{\top}}/{\sqrt{\vd_k}}\right) \mV
\end{equation}
with $\mQ=\mW_Q\mX$, $\mK=\mW_K\mX$, and $\mV=\mW_V\mX$. The entire mechanism can be intuitively understood by considering $\mQ$ and $\mK$ as the questions of the input sequence and the capability scores of every input token to answer each question. In return, the inner product $\mQ\mK^{\top}$ measures the weighted similarity of input tokens. Compared to recurrent networks, self-attention layers investigate pair-wise relationships along the entire chain to avoid squashing long-range dependencies in a vector. It then becomes theoretically possible to train extra-long sequences with little information loss. However, the massive size of learnable parameters limits the input sequence length to no more than $512$ or $1,024$ tokens in most models.

\subsection{Multimodal and Multiscale Molecular Representations}
While this review mainly discusses the graph representation of proteins, there is potential for integrating protein descriptions in various modalities and scales to uncover the mechanisms behind protein functionalities \cite{alber2019integrating}. 

A handful of previous literature features different morphologies of proteins, of which sequence and structure representations are arguably the two most popular options. Large pre-trained language models such as \textsc{ProtBERT} \cite{elnaggar2021prottrans} and \textsc{ESM-2} \cite{lin2022language} are used for sequential representations, while GNNs and CNNs can encode local environment information for structure representations. In downstream tasks, various algorithms such as MLPs \cite{guo2019accurate, zhang2019multimodal}, autoencoders \cite{tubiana2022scannet,WEI2022418}, or other specially-designed methods \cite{guo2019identifying, chen2022structure} can integrate separate embeddings from different modalities into a single-vector representation. Multiple sequence alignment (MSA) can also enhance sequential inputs to identify conserved and variable regions of proteins from homologous sequences \cite{jumper2021highly,meier2021esm1v,rao2021msa}. Other research seeks versatile modalities. 
\textsc{HOLOPROT} \cite{somnath2021multi} represents a protein's surface and structure, where the former generates the spherical relationship of residue nodes through a triangulation algorithm, and the latter captures four levels of a protein structure by a residue graph with multi-scale message-passing networks. \textsc{Transformer-M} \cite{luo2022one} establishes atomic-level descriptions for a protein by incorporating 1D (\eg physiochemical properties), 2D (\eg sequential distance), and 3D (\eg Euclidean distance) channels. Beyond conventional topological or physiochemical characteristics from sole proteins, some work explores feature representation from residue-centric meta-paths in interactive networks, such as PPI networks \cite{zhang2019multimodal,huang2022deepfusiongo}, drug-protein-disease networks \cite{xuan2021integrating} or combining six STRING networks for human and yeast \cite{gligorijevic2018deepnf}.

Describing proteins from different scales and obtaining more detailed representations while minimizing costs can be highly advantageous. Several methods have been proposed to extract feature representations for proteins and ligands at different levels. For example, \textsc{DeepFusion} \cite{song2022deepfusion} generates feature representations for drugs and proteins using pre-computed global structural similarity matrices and local chemical substructure semantic features. To encode structure representations at multiple scales, a coarse-grained GNN handles node features and edge connections at coarser levels with gated recurrent units \cite{liu2022multiscale}. \textsc{AGL-Score} \cite{nguyen2019agl} and \textsc{MWCG} \cite{rana2023geometric} construct multiscale weighted colored subgraphs with atomic types and extended connectivity interactive features (ECIF) to capture intramolecular interactions between proteins and ligands.
\textsc{iEdgeDTA} \cite{suviriyapaisal2023iedgedta} extracts both node-level and sequence-level protein features by language models and graph convolutions. \textsc{MSSA-Mixup} \cite{kong2023multi} captures internal dependencies among different scales of graph signals by graph wavelet theory using self-attention, and \cite{liu2023persistent} develops the persistent path-spectral model that characterizes the data at multiple scales during the filtration process.

\subsection{Model Augmentation}
The availability of accurately labeled biomolecular data is often limited, and the ability of a trained model to generalize to new data is often a major concern. \emph{Self-supervised learning} \cite{liu2021self, xie2022self} strategies are becoming increasingly popular as they offer a promising solution to these problems. These strategies train algorithms to recover partially corrupted or incomplete original inputs without any additional manual labels or supervision. Pre-trained models can then be directly applied to downstream tasks, and in situations where there is a small set of labeled training samples, self-supervised learning can replace random initialization, as seen in many computer vision tasks. Though, when fitness to a particular prediction task is the only concern and the number of labeled instances is sufficiently big, a supervised learning approach remains the preferred choice. This section introduces two broad categories of self-supervised learning strategies, which are generative and contrastive methods.

\subsubsection{Generative Learning Models}
Generative models aim at summarizing observed probability density landscapes in a hidden space and sample new molecules or molecular representations conditioned on protein properties, functions, and/or structures \cite{wu2021protein,strokach2022deep}.

\paragraph{Variational Autoencoders (VAE)} 
VAE refers to a general class of networks that map the inputs $\vx$ by an encoder network $q(\cdot)$ to a low-dimensional latent space that follows a specific distribution, \eg $\vz\sim\gN(0, I)$. A decoder network $p(\cdot)$ then reconstructs samples from the latent space to $\vx^{\prime}$ in the original space. The training goal is to find the optimal network parameters that minimize the distance between the reconstructed and original samples, \ie
\begin{equation} \label{eq:vae}
    \gL(p,q) = \operatorname{KL}\left( q(\vz|\vx)\|p(\vz) \right)+\lambda\gL_{\text{reconstruct}}(\vx,\vx^{\prime}),
\end{equation}
where $\lambda$ is a tunable parameter. The reconstruction loss $\gL_{\text{reconstruct}}$ accounts for inductive biases that are added to the inputs, such as the molecule's rigidity \cite{eguchi2022ig}.

\paragraph{Generative adversarial networks (GANs)}
GANs are a subset of energy-based models that comprise a \emph{generator} $G(\cdot)$ that maps random noise to instances in the data space and an \emph{adversarial discriminator} $D(\cdot)$ that distinguishes between real and generated examples. A min-max loss function pushes the generator and discriminator toward the dynamic equilibrium by
\begin{equation*}
    \gL(D,G)=\mathbb{E}_{\vx}\left[\log D(\vx)\right]+\mathbb{E}_{\vz}\left[\log(1-D(G(\vz)))\right],
\end{equation*}
where $\vx$ is sampled from ground truth input, and $\vz$ is generated using $G(\cdot)$. Maximizing the objective function with a fixed $G(\cdot)$ guides toward the optimal discriminator. There are various options for reformulating the discriminator loss, and a comprehensive discussion can be found in \cite{bond2021deep}.

\paragraph{Generative Diffusion Models} 
The explosive success in image synthesis has attracted increasing attention in generative diffusion models \cite{sohl2015deep,ho2020denoising,rombach2022high}. Similar to hierarchical VAE \cite{vahdat2020nvae}, a diffusion model steps into latent states progressively by $\vx\rightarrow\vz_1\rightarrow\dots\rightarrow\vz_T$. Compared to hierarchical VAE that learns an implicit transformation from $\vx$ to $\vz$, diffusion models add random Gaussian noise during the generation steps by $\vz_t=\vz_{t-1}+\epsilon$, $\epsilon\sim\gN$. Consequently, the loss function involves an additional noise matching term $\gL_{\text{noise match}} = \sum_{t=2}^T\mathbb{E}_{q(\vz_t|\vx)}\left[\operatorname{KL}\left( q(\vz_{t-1}|\vz_t,\vx)\|p(\vz_{t-1}|\vz_t) \right)\right]$ over Eq.~\eqref{eq:vae}. Generative diffusion models have been applied in several recent works for molecule generation or protein design, where noise is added to atoms' type \cite{corso2023diffdock} or structure \cite{xu2021geodiff,hoogeboom2022equivariant, wu2022protein}.

\subsubsection{Contrastive Learning}
Contrastive learning is another technique utilized to enhance the representation of proteins \cite{becker1992self, hadsell2006dimensionality}. Given a protein instance $\gG$ that is similar to $\gG_+$ and dissimilar to $\gG_-$, a function $f(\cdot)$ is used to encode the three graphs in order to maximize the similarity between $f(\gG)$ and $f(\gG_+)$ while minimizing the similarity between $f(\gG)$ and $f(\gG_-)$. To achieve this, pseudo data labels are required during self-supervised learning, where positive pairs ($\gG$ and $\gG_+$) can be jointly sampled by sequence-based \cite{hermosilla2022contrastive, wang2022molecular} or Euclidean distance-based \cite{zhang2022gearnet} croppings. Negative pairs ($\gG$ and $\gG_-$), on the other hand, are sampled independently. When measuring the pair-wise similarity of hidden representations, a scoring function is learned on the encoded representation. For instance, \textsc{InfoNCE} \cite{lu2020self,lu2022discovering,hermosilla2022contrastive} and \textsc{NT-Xent} \cite{wang2022molecular} evaluate the agreements between the positive pair $f(\gG),f(\gG_+)$ and all other negative pairs $f(\gG),f(\gG_-)$. Formally,
\begin{equation}
\label{eq:nce}
    \gL_{\text{NCE}} = -\mathbb{E}\left[\log\frac{\exp\left(\gS(f(\gG),f(\gG_+))\right)}{\exp\left(\gS(f(\gG),f(\gG_+))\right)+\sum_{i=1}^{B}\exp\left(\gS(f(\gG),f(\gG_{-,i}))\right)}\right]
\end{equation}
with batch size $B$. The similarity between two representations is usually implemented by the inner products $\gS(f(\gG),f(\gG_+))=f(\gG)^{\top}f(\gG_+)$ or the cosine similarity $\gS(f(\gG),f(\gG_+))=-\frac{f(\gG)^{\top}f(\gG_+)}{\|f(\gG)\|_2\|f(\gG_+)\|_2}$. The minimized loss attains a lower bound of the mutual information for the hidden representations. Moreover, researchers may consider integrating previous strategies, such as \textsc{MoHeG} \cite{liu2021representation}, which leveraged a weighted loss function that combined Eq.~\eqref{eq:vae} and Eq.~\eqref{eq:nce}. This approach led to a substantial improvement over using either equation individually.

\section{Applications}
\label{sec:application}
Proteins that possess functional activity are typically composed of multiple polypeptide chains or require the presence of other biomolecules, such as coenzymes and cofactors, to perform their biological roles. A number of routine tasks involving protein functionality have been investigated in the scientific literature. In accordance with Section~\ref{sec:4homoHeteroGraphs}, we categorize three levels of protein interactions with compounds, proteins, and biological networks, respectively. 

\begin{figure}[t]
    \centering
    \includegraphics[width=\textwidth]{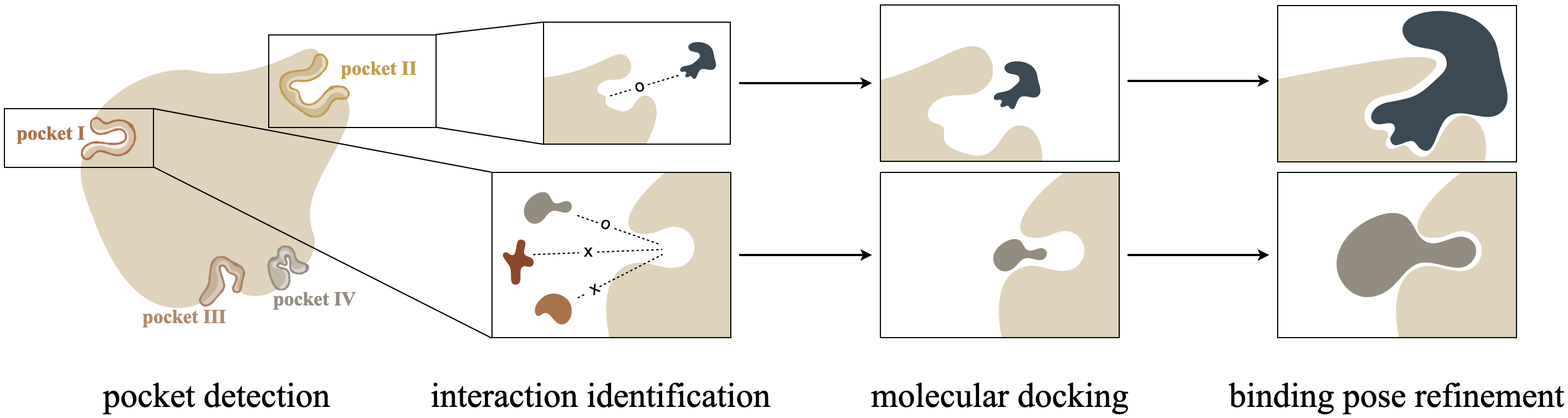}\vspace{-1cm}
    \caption{Four computational steps of a ligand binding to a particular receptor protein. }
    \label{fig:4stepsBinding}
\end{figure}

\subsection{Drug Discovery}
Drugs (ligands) exert their effects by interacting with target viral proteins (receptors). A ligand molecule binds specifically and reversibly to a receptor protein, such as a virus, to alter its activity or function. Therefore, designing a new drug involves identifying potential ligands from drug-like molecules and studying their interactions with specific receptors. Figure~\ref{fig:4stepsBinding} demonstrates the four computational steps for a ligand binding to a given receptor protein \cite{rachman2018predicting}: 1) \emph{pocket detection} finds feasible binding site(s) on the receptor. 2) \emph{interaction identification} recognizes candidate ligands that may interface with the current pocket. 3) \emph{molecular docking} generates binding poses for the ligand-receptor docking pairs. 4) \emph{binding pose refinement} dynamically fine-tunes the conformations at the binding sites.

\subsubsection{Pocket Detection}
Detecting new binding pockets is a crucial and challenging task that initiates the entire process of drug discovery. The identification of binding regions is typically achieved by locating heavy atoms in molecule sequences \cite{chen2022quotetarget,lee2022sequence} or voxels of 3D protein images \cite{aggarwal2021deeppocket,eguida2022target,gagliardi2022shrec} or surfaces \cite{mylonas2021deepsurf,gagliardi2022shrec,gagliardi2022siteferret}, where \textsc{Transformer} \cite{vaswani2017attention} or \textsc{3DCNN} \cite{torng20173d,lu2022machine} are frequently used encoding blocks. 

Since proteins do not naturally distribute atoms on a grid, recent work has explored using graph representations to capture protein structures. Binding site prediction is then converted into a classification task on graph nodes. \textsc{ScanNet} \cite{tubiana2022scannet} detects both protein-protein binding sites and B-cell epitopes (\ie antibody binding sites) to SARS-CoV-2 spike proteins with attention-based graph convolutions on both atom-level and residue-level protein representations. \textsc{GraphPPIS} \citep{yuan2021structure} extracts node representations of residue graphs with deep GCNs to identify PPI sites with protein-interacting probabilities. \textsc{PocketMiner} \cite{meller2022predicting} modifies \textsc{GVP-GNN} to find cryptic pockets of protein targets where binding sites are only exposed on the surface during binding. 

Another line of research focuses on designing \emph{pocket matching} algorithms to pair relative pockets for binding similar ligands, which can be assessed by the shape of molecular surfaces or sub-graphs that reveals similarities in their physical and biological properties \cite{kumar2018advances,naderi2019binding,simonovsky2020deeplytough}. In graph representation, the objective then turns to finding the maximum similarity between local sub-graphs, such as the pocket size, solvent accessibility, and flexibility \cite{govindaraj2018comparative, eguida2022estimating}. The topological characteristics of pockets are usually captured by PDEs \cite{zhao2018protein, manak2019voronoi}, which evolve a convex hull surface inward until it touches the protein surface everywhere. The algorithm not only estimates the surface area, volume, and depth of the pocket but also describes sub-pockets characterizations that determine multi-ligand synergistic interactions.

\subsubsection{Interaction Identification}
The identification of potential interactions between drugs and their targets is crucial for the development of safe and effective drugs \citep{bagherian2021machine,lindell2022multimodal}. Such interactions can involve various types of binding, activation, and inhibition. Researchers have designed different computational approaches that can predict interactions between compound-protein, protein-protein, and antibody-antigen.

\paragraph{Compound-Protein Interaction (CPI)}
The interaction of drug molecules with target proteins is essential for identifying hit and lead compounds  \cite{du2022compound} and recognizing potential drug side effects \cite{tsubaki2019compound}. Due to the limited availability of protein structures in the pre-AlphaFold era, earlier research encodes protein sequences by language models and represents compounds as sequences \cite{wang2020gnn,chen2020transformercpi,ma2022deep} or images \cite{qian2022picture}, or graphs \cite{elbasani2021gcrnn,qian2021spp, zhao2022cpgl}.  Later on, the increasingly available number of protein structures urges the prevalence of graph convolutions for protein receptors, such as GCN, \textsc{GraphSAGE}, and GAT, for CPI prediction. These approaches aim to predict binary interaction estimations \cite{wan2022inductive,knutson2022decoding} or numerical binding affinities\cite{feinberg2018potentialnet}.

 Predicting protein-protein interactions can be approached on different levels, such as atom, residue, or protein levels, for binding score prediction or binary interface label assignment. Graphs can also be constructed for interface regions for protein-protein complexes instead of individual proteins.

\paragraph{Protein-Protein Interaction (PPI)} 
In contrast to small molecules that interact with a single target receptor, proteins act in a network of complex molecular interactions, investigating which is vital to prevent side effects from false-binding \cite{lehne2009protein}. Predicting protein-protein interactions can be approached on different levels, such as atom \cite{jha2022prediction}, residue \cite{reau2023deeprank}, or protein \cite{mahbub2022egret} levels, for binding score prediction or binary interface label assignment. On top of individual proteins \cite{pancino2020graph}, the graph(s) might also be constructed for interface regions for protein-protein complexes \cite{reau2023deeprank}.

\paragraph{Antibody-Antigen Interaction} An antibody is a Y-shape protein of light and heavy chains. The variable domain, defined jointly on both chains, is a part of the antibody's binding region to recognize their respective antigens, \ie complementarity-determining regions (CDRs) \cite{ambrosetti2020modeling}. Overall, the target of antibody-antigen binding is similar to that of compound-protein interaction, where both tasks focus on analyzing the interface region (pocket or CDR region) to predict complex binding affinity and docking scoring function. In antibody-antigen interaction predictions, the interface regions might be described jointly by pharmacophores and atomic distance \cite{myung2022csm} or separately by residue graphs \cite{del2021neural,halfoncontactnet}.

\subsubsection{Molecular Docking}
Molecular docking finds the 3D structure of the ligand-receptor complex for optimal conformations, where many \textit{in silico} methods follow the `lock-and-key' theory \cite{morrison2006lock} and solve the correct orientation of the ligand to match the rigid receptor. To identify the tight pose of a protein-ligand complex, a docking program constructs a set of poses with minimized free energy and scores the strength of the putative poses to select the optimal pose for the top-rank docking pairs \cite{pantsar2018binding,zhu2022torchdrug}.

\paragraph{Binding Pose Generation}
Molecular docking predicts the best binding conformations for ligand-protein complexes with diverse optimized scores, such as the predicted error in binding affinity forecasting \cite{gao2022cosp}, the ligand position \cite{ganea2021independent}, or a combination of both measurements \cite{lu2022tankbind}. As a surrogate of ligand positions, generative models tend to target indirect predictors that are considered easier to learn, such as ligand position relative to the protein, orientation in the pocket, and the torsion angles \cite{corso2023diffdock}. Other attempts to enhance the prediction performance have been assessed by introducing trigonometry constraints for residue graphs \cite{jumper2021highly,jing2020learning} and reconstructing the future trajectories for the dynamic systems of complexes \cite{zhu2022neural}. Graphs are usually defined at the atomic level for drugs and the residue level for proteins, with encoders essential for describing the roles of key residues in the conformational transition \cite{zhu2022torchdrug,lu2022tankbind,stark2022equibind}.

\paragraph{Antibody-Antigen Docking and Design}
Antibody-antigen docking prediction falls into the general class of protein-protein docking and design, but the lack of evolutionary information in paratopes prevents many MSA-based protein design methods from predicting antibody structures. 
In addition to generating 3D coordinates for binding regions \cite{eguchi2022ig}, the highly variable contact regions also suggest designing the sequence of CDRs to complete the antibody-antigen design \cite{shin2021protein,akbar2022silico}. Generative models have been applied for such a problem \cite{jin2022antibody}, and the predictions have been tested on SARS-CoV-2 virus variants \cite{jin2021iterative,eguchi2022ig}.

\paragraph{Binding Affinity Prediction} 
The effectiveness of a particular drug stands on its affinity with the target protein receptors \cite{paul2021artificial}. Classic simulation-based approaches require the precise 3D conformation of drug-target pairs at the binding sites with hand-crafted physiochemical properties to calculate the affinity score. Deep learning methods \cite{shen2020machine, dhakal2022artificial, zhao2022brief}, instead, train on a large number of molecule pairs and fit the nonlinear relationship of the input sequential and structural features for predicting binding affinities. Pairs of molecules can either be learned independently in parallel, and then concatenated together in the hidden space \cite{karimi2019deepaffinity}, or established jointly in a single graph by taking, \eg the drug molecule as a node in the protein-based graph \cite{nguyen2021gefa} or the complex as a whole \cite{li2022multiphysical}. Furthermore, the examination of molecular dynamics (MD) trajectories during binding can enhance the accuracy of predicting binding affinities \cite{wu2022pre}.

\subsubsection{Binding Pose Refinement}
Protein binding sites are typically larger than the ligands they bind, and the binding pairs are considered rigid bodies during the docking process \cite{kahraman2007shape}. Hence, it is important to fine-tune the conformation of the complex by adjusting the bound pose \cite{nguyen2021gefa,jiang2022predicting,wankowicz2022ligand}. Updating the binding pose of the input ligand involves iterative progress when predicting the conformations of docked molecules. Nonetheless, this process of generating 3D atomic coordinates of the conformation can introduce errors that may bias the final result \cite{mansimov2019molecular,zhang2022e3bind,xu2021geodiff,mansimov2019molecular} can accumulate errors, leading to bias in the final result. To mitigate this issue, the predictor(s) may adopt an approach that characterizes intermediate geometric elements to maintain roto-translation invariance, such as atomic distance \cite{kohler2020equivariant,simm2020generative,xu2020learning} and torsion angle \cite{gogineni2020torsionnet,ganea2021geomol}.

\subsection{Protein Understanding}
Investigating proteins and protein-associated interactions is essential for understanding the mechanism of many reactions in cytology. This section digests two specific applications, \ie nucleic-acid-binding residues and polymers.

\subsubsection{Nucleic-Acid-Binding Residues} 
RNA-binding proteins (RBPs) interact with RNAs through RNA-binding domains (RBDs) for chemical synthesis to regulate RNA metabolism and function. Conversely, RNAs can bind to RBPs to modulate the fate or activity of the bound RNAs \cite{hentze2018brave}. Due to the heterogeneity of RBDs and the existence of unconventional RBPs that lack identifiable RBDs, geometric learning methods are effective options in discovering new binding sites \cite{uhl2021graphprot2} and identifying binding residues \cite{xia2021graphbind}. 

Protein-RNA interactions are essential in cellular processes, such as gene expression regulation, protein synthesis, and post-transcriptional processes in all eukaryotes. The interaction between non-coding RNAs (ncRNAs) and proteins (ncRPIs or NPIs) is particularly critical in gene regulation and human diseases \cite{esteller2011non,zhang2017computational}, and they can be formulated similarly to other molecule interaction prediction tasks as typical binary classifications on graph pairs \cite{wang2021edlmfc}. More work views it as a link prediction task on a bipartite graph with both ncRNAs and proteins being nodes featured with embedded sequence or structure features \cite{ge2016bipartite, nacher2019controllability,shen2021npi,philip2021survey}. Notably, such a graph only connects nodes with firm interactions, \ie it does not include negative labels on edges. Moreover, predicting ncRPIs is generally cell-specific, necessitating additional techniques like transfer learning \cite{arora2022novo} to accommodate interactions in different cell lines or other diverse contexts.

In addition to RNA, the investigation of protein interactions with DNA is another important area of research that plays a key role in cellular functions such as transcriptional regulation, chromosome maintenance, and replication \cite{dey2012dna}. Although DNA and RNA have distinct biological functions, they share similar components \cite{tajmir2006overview} and structure \cite{reichl2015understanding, miyahara2016similarities}. Thus, the RPI models developed for RNA-protein interactions can potentially be adapted for studying DNA-protein interactions \cite{yuan2022alphafold2,zhang2022deepdisobind}. Recent studies have facilitated a more in-depth understanding of the complex mechanisms underlying DNA-protein interactions.

\subsubsection{Multimeric Protein Designs}
Multimeric proteins are composed of two or more connected polypeptide chains. Deep learning techniques have facilitated the prediction and optimization of novel polymers \cite{sha2021machine,li2023neural}. State-of-the-art protein structure prediction models such as \textsc{AlphaFold 2} can fold proteins with multiple chains; however, the prediction accuracy decreases as the number of chains increases \citep{bryant2022predicting}. This necessitates adjustments to account for the difference between folding single-chain and multimeric proteins. One approach is to define graph representations for protein complex structures that include a binary attribute on edges, indicating whether two neighboring nodes come from the same chain \cite{chen2022dproq,morehead2022egr}. Additionally, modifying the loss function enables capturing inter-chain interfaces \cite{dauparas2022proteinMPNN} and respecting permutation symmetry in multi-chain alignment \cite{evans2022multimer}. Recent studies have demonstrated the effectiveness of these approaches, emphasizing the significance of taking into account the intricate interplay among multiple chains in multimeric proteins.

\subsection{Biological System Analysis}
Networks are useful for comprehending and forecasting intricate biological systems by providing linkage maps that connect various entities, including genotypes, phenotypes, and environmental factors. As one of the advanced system modeling and simulation techniques, GNNs reveal the hidden effects of diverse factors on biological entities of interest through analyzing the heterogeneous graph \cite{wu2021hashing}.

\subsubsection{Metabolic Network}
Metabolic networks consist of complex enzymatic reaction pathways that convert nutrients into energy required for biological processes. Due to the intricate relationship between enzymes and reactions, these networks are naturally suitable for network analysis beyond pairwise connections. Furthermore, since the direction of reaction flows can be altered depending on the environmental context, GNNs designed for metabolic networks employ directional message aggregations to examine flux balance.

Although no standard paradigms have been established, a few researchers have made strides in understanding metabolic systems using graph representations. A cell-level directed factor graph learns to estimate metabolic stress, recover noisy metabolic genes, and segment cells from the substrate or product states of metabolites in reactions \cite{alghamdi2021graph}. The importance of directionality and environmental perturbations has been verified for capturing the flow of metabolites produced or consumed by enzymatic reactions \cite{beguerisse2018flux}. Furthermore, by identifying associations between metabolite-disease, metabolite-metabolite, and disease-disease, potential metabolite-disease relationships were inferred \cite{sun2022deep}.

\subsubsection{Gene Regulatory Networks}
Gene Regulatory Networks (GRNs) uncover co-expression patterns among regulators and target genes. Validating and inferring (potential) regulatory associations are preliminary for identifying disease pathways or novel therapeutic targets, which essentially answers how functional development is controlled by heritable genomic sequence information \cite{levine2005gene}. 

For GRN-related learning tasks, a heterogeneous graph is often used to analyze the intrinsic structure-function relationship. The nodes in such graphs can represent various biological units, such as multi-omics data (\eg mRNA expression, microRNA expression, and DNA mutation)\cite{liu2021representation}, transcription factors \cite{wang2020inductive}, and genes \cite{patil2022learning}. Different characterizations of nodes and edges can lead to diverse learning tasks, such as interpreting the effects of functional interactions  \cite{liu2021representation}, uncovering connections between transcription factors or TF-gene pairs \cite{wang2020inductive}, and learning causality between temporal elements in dynamic GRNs \cite{patil2022learning}.

\subsection{Polypharmacy and Drug Repurposing}
Pharmaceutical companies often repurpose existing drugs for new diseases as a cost-effective strategy. The optimal drug molecules are those that have the highest negative binding energy, which leads to the strongest binding affinity between ligands and the target protein. In contrast, for non-target proteins, a preference for weak binding is observed to avoid misbinding and associated side effects \cite{pan2022deep,paul2021artificial}.

To study drug repurposing, a heterogeneous graph can be generated with nodes representing drug molecules, diseases, proteins, and other relevant objects. The primary learning objective is typically a link prediction task, where connected edges indicate the positive or negative effects between drug-disease pairs \cite{doshi2022computational} or synergistic effects of drug combinations \cite{jiang2020deep}. For a more accurate inference, extra information such as gene-related embeddings \cite{mei2022relation} and patient information from medicines knowledge graphs \cite{shang2019gamenet,wang2019order} can be incorporated.

In addition to node types, it is important to consider different edge types that describe the various interfaces between drugs, diseases, and other entities. This is particularly relevant for recognizing polypharmacy side effects, where the interactions between multiple drugs can lead to unexpected adverse effects \cite{wang2022extending,manoochehri2019graph,zitnik2018modeling}. Various methods have been proposed to build these complex interactions, including extending edge types in heterogeneous graphs, graph-based machine learning models, and network analysis approaches.

\subsection{Others}
Recent developments in \emph{single-cell RNA-sequencing} (scRNA-Seq) have enabled the simultaneous measurement of RNA and protein abundances at the single-cell level. This has provided a unique opportunity to investigate the heterogeneity and dynamics of tissues, organisms, and complex diseases. The use of graph representation has facilitated the understanding of various biological processes, including protein interactions \cite{dai2021pike}, gene expression \cite{li2022improving}, and cell-cell relationships \cite{wang2021scgnn}.

GNNs have proven to be versatile tools for addressing various problems in biological research, including \emph{genetic mutations} and \emph{automated laboratory results}. For instance, \textsc{VGEN} \cite{cheng2021vegn} utilizes a heterogeneous graph of genes and mutants to model variant effects prediction and identify disease-causing mutations from millions of genetic mutations in an individual patient. Meanwhile, \textsc{ModelAngelo} \cite{jamali2022modelangelo} leverages GNNs and hidden Markov models to map protein chains to entries in a user-provided sequence file. The model refines residue-based graph representation with additional data sources, such as cryo-EM data, amino acid sequence data, and prior knowledge about protein geometries.

\section{Benchmarks Datasets}
\label{sec:benchmark}
Although deep learning methods exhibit promising performance in many complex biological problems, building a reliable model necessitates having sufficient high-quality data. In this regard, we have reviewed $11$ popular databases that can be classified into four categories: protein complexes, protein-ligand bindings, antibody-antigen structures, and knowledge graphs. The first category contains protein structures that may not necessarily include target complexes for prediction but can aid algorithms in understanding the semantics and syntax of proteins. Next, several databases have been established for protein-protein, drug-target, and antibody-antigen binding pairs, with learning tasks that can be regression or classification. For drug repurposing knowledge graphs, we provide an example dataset of about $100,000$ entities. Nevertheless, the majority of research works prefer to construct their knowledge graphs as data preparation is a key step, and the analyzed datasets are usually case-specific. 

\begin{table}[H]
\caption{Benchmark datasets for interactive protein learning tasks.}
\vspace{-5mm}
\label{tab:dataset}
\begin{center}
\resizebox{\linewidth}{!}{
\begin{tabular}{ccccl}
\toprule
\textbf{Type} & \textbf{Dataset}(ver) &\textbf{Size} & \textbf{Learning Task}$\dagger$ & \textbf{Summary} \\
\midrule
\multirow{7}{*}{\rotatebox[origin=c]{90}{Protein Complexes}} 
& \makecell{\textbf{PDB} \cite{wwpdb2019protein} \\ (2023.01)} & \makecell{$215,100$\\proteins} & NR \cite{jumper2021highly,evans2022multimer} & \makecell[l]{proteins and nucleic acids complexes (with small molecules)\\and related metadata (\eg experimental results).\\
\url{https://wwpdb.org/}} \\
& \makecell{\textbf{CATH-Plus} \cite{sillitoe2021cath}\\v4.3} & \makecell{$500,238$\\protein domains} & NC \cite{jing2020learning,dauparas2022proteinMPNN} & \makecell[l]{non-redundant tertiary structure of protein domains and\\super-families of protein domains from \textbf{PDB}.\\
\url{https://www.cathdb.info/}} \\
& \textbf{AlphaFold DB} \cite{tunyasuvunakool2021highly} & \makecell{$200$M\\proteins} & \cite{jendrusch2021alphadesign} & \makecell[l]{A massive database covering $98.5\%$ of human proteins folded\\from protein sequences by \textsc{AlphaFold 2}.\\
\url{https://alphafold.ebi.ac.uk/}} \\
\midrule
\multirow{18}{*}{\rotatebox[origin=1B]{90}{Protein-Ligand Binding}} & \makecell{\textbf{PDBbind-CN}\cite{liu2017forging}\\(2020)}& \makecell{$23,496$\\complexes} & \makecell{GR \cite{feinberg2018potentialnet,nguyen2021gefa,li2022multiphysical,jiang2022predicting, knutson2022decoding},\\NR \cite{stark2022equibind,lu2022tankbind,jiang2022predicting,zhang2022e3bind,corso2023diffdock}} & 
\makecell[l]{complexes with experimentally measured binding affinities\\on protein-ligand (19,443), protein-protein (2,852), protein\\-nucleic acid (1,052), and nucleic acid-ligand (149) complexes.\\
\url{http://www.pdbbind.org.cn/}}\\
& \makecell{\textbf{BindingDB} \cite{liu2007bindingdb}\\(2022.11)} & \makecell{$2,613,813$\\pairs} & GR \cite{karimi2019deepaffinity} & \makecell[l]{binding affinity between proteins ($8,942$) considered to be\\candidate drug-targets and ligands that are small, drug-like\\molecules ($1,123,939$).\\\url{https://www.bindingdb.org/rwd/bind/index.jsp}} \\
& \textbf{DB5.5} \cite{guest2021expanded} & \makecell{$257$\\pairs}	& \makecell{GR \cite{reau2023deeprank},\\NC \cite{del2021neural},\\NR \cite{ganea2021independent}} & \makecell[l]{non-redundant protein–protein complexes with the unbound\\structures of their components from \textbf{PDB}.\\\url{https://zlab.umassmed.edu/benchmark/}} \\
& \makecell{\textbf{Dockground} \cite{collins2022dockground}\\(2023.06)} & \makecell{$667,331$\\interfaces\\$363,240$\\protein chains} & \makecell{NC \cite{tubiana2022scannet}\\GR \cite{myung2022csm}}
& \makecell[l]{a comprehensive database of experimentally determined and\\simulated protein structures, model-model complexes, docking\\decoys of experimentally determined and modeled proteins,\\and templates for comparative docking.\\\url{https://dockground.compbio.ku.edu/}} \\
& \makecell{\textbf{ZDOCK} \cite{hwang2010protein}\\v4} & \makecell{$176$\\unbound protein pairs} & GR \cite{myung2022csm} & \makecell[l]{high-resolution protein–protein complex structures by X-ray\\or NMR unbound structures, including $52$ enzyme-inhibitor,\\$25$ antibody–antigen, and $99$ complexes with other functions.\\\url{https://zlab.umassmed.edu/zdock/benchmark.shtml}}\\
\midrule
\multirow{2}{*}{\rotatebox[origin=c]{90}{Antibody-Antigen }} & \textbf{AbDb} \cite{ferdous2018abdb} & \makecell{$5,976$\\antibodies} & \makecell{GC \cite{halfoncontactnet},\\NR \cite{eguchi2022ig},\\NC \cite{del2021neural}}
& \makecell[l]{antibody structures summarized from \textbf{PDB} by \textbf{SACS}.\\\url{http://www.abybank.org/abdb/}}\\
& \makecell{\textbf{SAbDab} \cite{schneider2022sabdab}\\(2023.03)} & \makecell{$7,073$ antibody\\$5,544$ antigen} & \makecell{NC \cite{tubiana2022scannet,del2021neural},\\NR \cite{jin2021iterative,jin2022antibody}}
& \makecell[l]{all the antibody structures available in \textbf{PDB}. \\\url{https://opig.stats.ox.ac.uk/webapps/newsabdab/sabdab/}}\\
\midrule
KG & \textbf{DRKG} \cite{drkg2020} & \makecell{$97,238$\\entities} & LC \cite{doshi2022computational} & \makecell[l]{$107$ types of relationships among $13$ types of entities (genes,\\compounds, diseases, biological processes, side effects, and\\symptoms) from various sources. \\\url{https://github.com/gnn4dr/DRKG/}}\\
\bottomrule
\multicolumn{5}{l}{\makecell[l]{$\dagger$ Learning tasks are categorized into node (N), graph (G), and edge (L) level classification (C) or regression (R),\\ \eg, NC denotes node classification.}
}
\end{tabular}
}
\end{center}
\end{table}

We summarize the $11$ databases and link them to commonly used learning tasks. Depending on the database and research question, predictors can be defined on the node, edge, or graph level, with predicted labels that are either continuous (\eg 3D coordinate) or discrete (\eg amino acid type), which we designate as regression or classification, respectively.

\section{Discussion}
\label{sec:discussion}

The scientific discovery field has seen the entrance of an increasing number of big tech companies that employ artificial intelligence, such as Alphabet, Meta AI, Microsoft, and Byte Dance. Alongside these companies, unicorn firms like AI Protein, SomaLogic, and ProteinQure use deep learning techniques to unlock the potential of protein-based therapeutics, and small startups offer \textit{in silico} computational experiments tailored to specific client needs.
Moreover, the open-sourcing of \textsc{AlphaFold 2} in 2021 has enabled accurate prediction of protein tertiary structures, which has greatly enriched available protein datasets and facilitated algorithm development. Despite the many advances in protein design and drug discovery, significant gaps still exist in computer science, biology, and production, and there is a high demand for developing new AI models for downstream tasks that remain unmet.

The limitation of current models poses a significant challenge in various applications. The main issue with deep learning models is their dependency on a substantial amount of data, yet the availability of qualified samples in scientific research is either under-discovered or extremely scarce. Moreover, it is crucial to develop models that can provide interpretable results for scientific problems, especially those focused on understanding the mechanisms of complex systems. However, the development of effective AI tools for dynamic mechanisms, such as protein folding and inverse folding, remains an open challenge. Another concern is those current AI methods seldom practice prior knowledge that was induced by biologists, which not only leads to inefficient use of expertise and widens the knowledge gap between computer science and biology, but also results in unnecessary computational resource consumption.

On the other hand, the lack of interdisciplinary communication poses a challenge for deep learning models in meeting scientific demands, leading to impractical optimization targets or evaluation rules for established problems \cite{tripp2022evaluation}. As a result, computational methods tend to focus on a narrow range of research topics, leaving many areas untouched or poorly resolved. While this review provides an incomplete overview of protein-involved interaction tasks explored by geometric deep learning methods, there are still many tasks without contemporary solutions. For instance, in protein binding systems, matchmaker \cite{sancar1993,meredith2016rna,chen2022working}, hub proteins \cite{tsai2009protein,vandereyken2018hub,jespersen2020emerging}, and metal ion–protein \cite{UEDA20031,grasso2013plasmonics,lin2022mass} bind partners to execute biological and physiological processes, but existing solutions lack modern deep learning methods. Developing new AI models that can address these deficiencies is a critical research direction for the future.

\bibliography{0main.bbl}

\end{document}